\documentclass[floatfix,twocolumn]{revtex4}
\usepackage{graphicx}
\usepackage{dcolumn}
\usepackage{colordvi}
\usepackage{color}
\usepackage{ulem}

\begin{document}

\title{Edge states and topological properties of electrons on the bismuth on silicon surface with giant spin-orbit coupling}

\author{D.V. Khomitsky}
\email{khomitsky@phys.unn.ru}
\author{A.A. Chubanov}
\affiliation{Department of Physics, University of Nizhny Novgorod, 603950 Gagarin Avenue 23, Nizhny Novgorod, Russian Federation}

\begin{abstract}

We derive a model of localized edge states in the finite width strip for two-dimensional electron gas
formed in the hybrid system of bismuth monolayer deposited on the silicon interface and described by the nearly-free electron model
with giant spin-orbit splitting. The edge states have the energy dispersion in
the bulk energy gap with the Dirac-like linear dependence on the quasimomentum and the spin polarization coupled
to the direction of propagation, demonstrating the properties of topological insulator.
The topological stability of edge states is confirmed by the calculations of
the $Z_2$ invariant taken from the structure of the Pfaffian for the time reversal operator for the filled bulk bands in the surface Brillouin zone
which is shown to have a stable number of zeros with the variations of material parameters.
The proposed properties of the edge states may support  future advances in experimental and technological applications of this new material
in nanoelectronics and spintronics.

\pacs{73.20.At, 75.70.Tj, 85.75.-d}

\end{abstract}

\maketitle

\section{Introduction}

During the last decade an increasing attention is given to a new class of structures called topological insulators (TI) with promising characteristics
both in fundamental aspects of their physics and possible applications in nanoelectronics, spintronics, and fabrication of new magnetic,
optical and information processing devices.\cite{KaneRMP,QiZhangRMP,Moore2010,Culcer,Qi2008}
The principal features of TI include the presence of time-reversal (TR) invariance in the system where the propagating edge states may exist being localized near
the boundary of host material and have the dispersion relation which is linear near the origin of their quasimomentum (Dirac-like structure),
corresponding to the energies belonging to the insulating gap of the bulk material.
The spin of such states is firmly attached to the direction of propagation along the
edge, making them protected against backscattering due to the TR invariance which
leads to effective cancelling of two scattered states with opposite possible directions of
the spin flip which accompanies such backscattering.
The existence of such edge states have been shown
in numerous theoretical models of TI, and also in the experiments. The materials included graphene,\cite{KaneMele}
HgTe/CdTe quantum wells,\cite{BHZ,Koenig,Zhou,Krueckl2011} bismuth thin films,\cite{Murakami} quantum wires\cite{Huber2011}, nanocontacts or bilayers,\cite{Sabater2013}
the  LiAuSe and KHgSb compounds,\cite{ZhangChadov}
as well as general two-dimensional models of paramagnetic semiconductors\cite{QiWuZhang}, the silicene,\cite{Ezawa,Drummond2012}
and the topological nodal semimetals.\cite{Burkov2011}
Another 2D TI has been predicted in the inverted Type II semiconductor InAs/GaSb quantum well,\cite{Liu2008}
and observed experimentally in the contribution of the edge modes to the electron transport.\cite{Knez2011}
Also a lot of studies has been devoted to the general properties of two- and three-dimensional models of TI with
certain symmetries,\cite{FuKane,FuKaneMele,Fukui2007,Fu,Drummond2012,Hou2011,Levin2011,Li2012}
where four topological invariants have been found in 3D TI instead of the single $Z_2$ invariant
in 2D TI.\cite{FuKaneMele,KaneRMP,QiZhangRMP}

Recently a general group theoretical analysis has been made for the links
 between the geometry of the Bravais lattice and the properties of TI.\cite{Slager}
It should be mentioned that the symmetry arguments have always played a significant role in classifying the systems as trivial or
topologically protected against external perturbations.\cite{KaneMele,FuKane,Slager,Doucot,Schnyder2008,Mirlin2010}
The time-reversal property of spin-1/2 particles in such systems can be described by the presence of time reversal invariance
(i.e. without the magnetic impurities or external magnetic field) and the absence of the spin rotational invariance,
here the time reversal operator is given by $\Theta=i \sigma_y K$, where $K$ is the complex conjugation operator
and $\sigma_y$ is the second Pauli matrix. According to the general symmetry considerations,\cite{Schnyder2008,Mirlin2010}
this means the class AII symmetry for the Hamiltonian for which the so called $Z_2$ topological order is possible for
two- and three-dimensional systems, forming the basis for the TI properties.

The studies of three-dimensional materials were mostly focused on the $\text{Bi}_2\text{Se}_3$,
$\text{Bi}_2\text{Te}_3$ or $\text{Bi}_2\text{Te}_2\text{Se}$ \cite{KaneRMP,QiZhangRMP,Hsieh2008,Miyamoto2012,Nechaev2013}
where also the edge states were constructed explicitly in several models of the finite-size geometry.\cite{Linder,Lu}
Another important issue is the effects of impurities and disorder on the band structure and topological stability in TI.
It is known that TI are robust against weak disorder or the potentials produces by non-magnetic impurities,
\cite{KaneMele,FuKane,ShengHaldane2006} while the presence of magnetic impurities may lead to the hybridization
of the insulator atomic orbitals and the magnetic material orbitals, producing the strong modification to metallic or non-metallic
nature of the states and their spin polarization. \cite{Caprara2012} Even for non-magnetic impurities it has been shown recently that the formation of impurity
bands within the energy gap at strong doping of the bulk material may lead to their mixing with the edge states of TI, modifying their structure, however keeping
the $Z_2$ order and the topological stability.\cite{Lee2013}

It can be seen that although the features of the TI are very general and describe a truly novel state of matter, the number of different materials demonstrating these
features is currently  quite limited. So, it is of interest to find new materials and compounds where possible manifestations of TI may be present, for both fundamental
aspects and applied purposes. It is needful also to understand which properties of the edge states are common for different systems, and which are special, and how all
of them are related to the bulk   quantum states in a specific model.

Here we consider a model of edge states and connect their properties to the topological characteristics of the host material for a new candidate for the class of
topological insulators: the two-dimensional electron gas (2DEG) in a material with strong spin-orbit coupling (SOC) formed at the interface of the monolayer of bismuth deposited
on the silicon. This material is characterized by a giant SOC splitting which was predicted or observed experimentally also in a number of metal films or the combined materials of
the "metal on semiconductor" type, \cite{Hirahara2006,Hirahara2007,Grioni2008,Dil2008,Gierz2009,Azpiroz2011,FPG} and recently described theoretically.\cite{FPG,JETP} Its huge spin splitting
together with the hexagonal type of the lattice creates a certain potential of manifestation of the TI properties, since the spin-resolved bands may evolve into spin-resolved
edge states, and the hexagonal type of the lattice is favorable for the TI to exist.\cite{Slager}

The properties of 2DEG at the Bi/Si interface have been studied experimentally with the help of angle-resolved photoemission
spectroscopy (ARPES) \cite{FPG,Hirahara2006,Hirahara2007,Hirahara2007b,Grioni2008,Dil2008,Gierz2009,Bian2009,Sakamoto2009} applied also to other materials.
It was found that this material represents an example of nowadays widely studied class of materials with large (up to $0.2 \ldots 0.4$ eV) SOC spin splitting of their energy bands,
which can be formed in various compound materials or heterostructures of the "metal on semiconductor" type.
It is known for many years that SOC plays an important role in formation of the TI properties,\cite{Sheng1996}
including the localization effects of Rashba SOC combined with electron-electron interaction,\cite{Strom2010}
the Dirac-cone surface states in $\text{Bi}_2\text{Se}_3$\cite{Basak2011}  and $\text{Bi}_2\text{Te}_x\text{Se}_{3-x}$,\cite{Niesner2012}
$\text{Pb}\text{Sb}_2\text{Te}_4$ or $\text{Pb}_2\text{Bi}_2\text{Te}_2\text{S}_3$,\cite{Protogenov}
and $\text{Bi}_{1-x}\text{Sb}_x$,\cite{Zhang2009} the topological phases\cite{Chern2010,Ruegg2012}
and quantum spin Hall phase in honeycomb lattice,\cite{Bercioux2011} the ultracold Fermi gases,\cite{Sau2011}
the spin Hall effect in graphene,\cite{Dyrdal2012} and the Kondo insulator effects.\cite{Dzero2012,Craco2012}

Various materials with strong SOC have been a subject of intensive studies throughout recent years, including the structures of Bi deposited on Si-Ge substrates,\cite{Miwa2005}
the Pb on Si structure,\cite{Dil2008} the trilayer Bi-Ag-Si system,\cite{Grioni2008}  the structures with monolayer of Pb atoms covering the Ge surface,\cite{Yaji2010}
or the Pb on Ge structures.\cite{Sherman2012} One can mention also new types of triple bulk compounds with strong SOC like $\text{Ge} \text{Bi}_{2} \text{Te}_4$,\cite{Okamoto}
BiTeI or other bismuth tellurohalides,\cite{Ishizaka2011,Gnezdilov2013,Chulkov}
 or recently discussed $\text{Bi}_{14} \text{Rh}_{3} \text{I}_9$ material.\cite{Rasche2013}

In the present paper we adopt the nearly-free model of two-dimensional bulk states in Bi/Si developed earlier\cite{FPG} and applied in the extended form in our previous paper
for the description of spin polarization, charge conductance and optical properties of this promising material\cite{JETP} for the calculation of the 1D edge states of
the electrons on the Bi/Si interface in the finite strip geometry. We obtain both the explicit form of the edge wavefunctions and the edge energy spectrum,
calculate their spin polarization, and link the possible topological stability of their properties to the $Z_2$ topological invariant studied by the analysis of
the time reversal matrix elements behavior in the Brillouin zone.\cite{KaneRMP,QiZhangRMP,Qi2008,KaneMele,FuKane,Fukui2007}
The results of our paper are of interest for expanding the knowledge of new materials with the topologically protected properties
where the SOC plays a significant role, making them suitable for further applications in spintronics as stable current-carrying and spin-carrying channels.

The paper is organized as following. In Sec.II we briefly described the nearly free-electron (NFE) model of 2D bulk states at Bi/Si interface, and derive a model for
the 1D edge states for the electrons in the finite strip geometry. We calculate their spectrum, wavefunctions,
and spin polarization. In Sec.III we reinforce our findings on the edge state stability by considering
 the topological band properties  of the 2D bulk states in Bi/Si, and find the results supporting the presence of the TI phase. Our conclusions are given is Sec.IV.

\section{Model for bulk states and edge states}

\subsection{Bulk states}

Our model for the 1D edge states is based on the 2D NFE model for the bulk states of 2DEG formed at the interface of the
trimer Bi/Si(111) structure\cite{FPG} developed for the description of the spectrum near the M-point of the BZ, and later extended for the modeling
of the electron states in the whole BZ.\cite{JETP} This model was compared with its expansion containing the anisotropic terms in the NFE model as well as with an empirical
tight-binding model.\cite{FPG} While the details of band structure and the quality of reproduction of experimental ARPES data on energy bands in Bi/Si vary from model to model,
the simple NFE model allows to reconstruct the main properties of spin split bands
including the magnitude of splitting, the energy gap, and the spin polarization. It also has a major advantage of a straightforward derivation
of the edge states in a finite strip geometry, as we shall see below.

In this model, the Hamiltonian for the 2DEG in the BZ of $(k_x,k_y)$ plane
$H=H_0+V(x,y)$ is written as a sum of free-electron term with SOC,

\begin{equation}
H_0=\frac{\hbar^2 k^2}{2m}+\alpha_R(\sigma_x k_y - \sigma_y k_x),
\label{h0}
\end{equation}

$k^2=k_x^2+k_y^2$, corresponding to the Rashba paraboloid
centered in the Gamma point of the hexagonal BZ, and the lattice potential represented via the spatial Fourier
expansion with reciprocal space vectors ${\bf G}_i$ as

\begin{equation}
V(x,y)=\sum_i V_i e^{i{\bf G}_i \cdot {\bf r}}.
\label{vxy}
\end{equation}

\begin{figure}[tbp]
\centering
\includegraphics*[width=85mm]{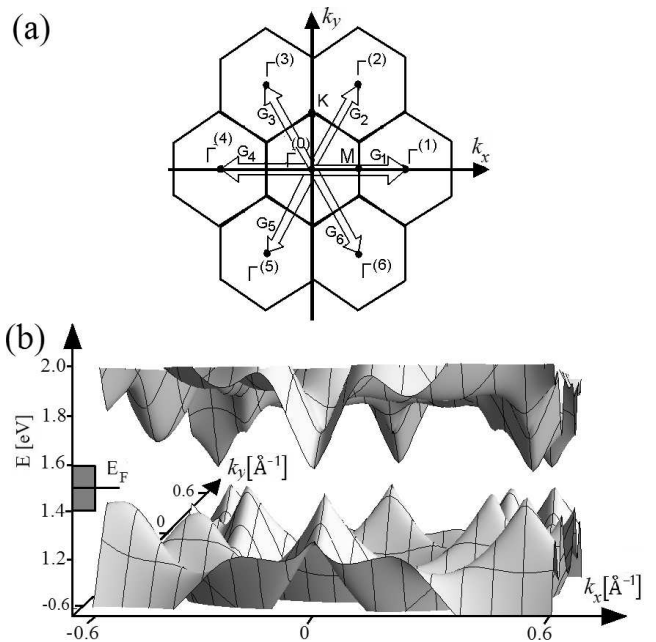}
\caption{(a) The structure of reciprocal space vectors connecting the equivalent Gamma points
$\Gamma_0, \ldots, \Gamma_6$ in the hexagonal lattice for the NFE model.
The parameters are $\Gamma M=0.54$ ${\AA}^{-1}$ and $\Gamma K=0.62$ ${\AA}^{-1}$.
(b) Fragment of the energy spectrum of 2DEG at the Bi/Si(111) surface in the NFE model
corresponding to the bulk band gap (marked by $E_F$ on the left part of the figure) of around $0.2$ eV between the bands
where the Fermi level is located, creating
the possibility of the edge state existence within this gap, and making the system a possible new topological insulator.}
\label{figbulk}
\end{figure}

The parameters of both
the $H_0$ and $V$ are fitted as to provide the best correspondence between the model and the structure of the bands near the Fermi level experimentally known
from ARPES measurements.\cite{Grioni2008,Dil2008,Gierz2009,FPG}
The typical values\cite{FPG,JETP} are $m=0.8 m_0$, $\alpha_R=1.1$ $\rm{eV \cdot \AA}$ and $V_i=V_0=0.3$ eV,
although they should be treated as fitting parameters rather than measured material constants, and in the present paper we shall
consider their variations in the range of $0.3 \ldots 0.6$ eV for $V_0$ and $0.6 \ldots 1.1$ $\rm{eV \cdot \AA}$ for $\alpha_R$.
The structure of reciprocal space vectors connecting the equivalent Gamma points $\Gamma_0, \ldots, \Gamma_6$
in the hexagonal lattice is shown in Fig.\ref{figbulk}a. In the model originally proposed\cite{FPG} only the vectors
${\bf G}_1, {\bf G}_2, {\bf G}_6$ were included in order to described the states near the M point, and later we expanded
this model\cite{JETP} with vectors ${\bf G}_3, {\bf G}_4, {\bf G}_5$ for the description of the states in the whole BZ.
The parameters of the hexagonal lattice in Fig.\ref{figbulk}a are $\Gamma M=0.54$ ${\AA}^{-1}$ and
$\Gamma K=0.62$ ${\AA}^{-1}$.

The Bloch eigenstate of the Hamiltonian $H=H_0+V$ is a two-component spinor which can be constructed under
the NFE approximation in the following form:

\begin{equation}
\Psi^{bulk}_{{\bf k}}({\bf r})=\sum_n a_{n{\bf k}} \psi^{bulk}_{n{\bf k}}({\bf r})
\label{bulkwf}
\end{equation}

where the Rashba eigenstates $\psi^{bulk}_{n{\bf k}} = \psi^{bulk}_{\bf{k}+\bf{G}_n}$ have the form of free electron states with
the quasimomentum shifted by the vector $\bf{G}_n$,
see Fig.\ref{figbulk}a, and

\begin{equation}
\psi^{bulk}_{{\bf k}}=\frac{e^{i {\bf kr}}}{\sqrt{2}}
\left(
\begin{array}{c}
1 \\
\pm e^{i {\rm Arg} (k_y-i k_x)}
\end{array}
\right).
\label{rf}
\end{equation}

The $(\pm)$ sign corresponds to two eigenvalues for Rashba energy spectrum
$E(k)=\hbar^2 k^2/2m \pm \alpha_R k$.

In Fig.\ref{figbulk}b we plot the fragment of the energy spectrum of 2DEG at the Bi/Si(111) surface in the NFE model
corresponding to the bulk band gap (labeled by $E_F$ at the left part of the figure) of around $0.2$ eV between the bands
where the Fermi level is located, reported to be above the second spin split band.\cite{FPG}
The Fermi level position in the bulk gap where the gapped structure of the electron spectrum is produced
by the hexagonal lattice with potential (\ref{vxy}) creates the possibility of the edge states formation with energies in this gap,
and, as we shall see below, makes the system a possible new candidate for
the topological insulator class. The large metallic-like values of electron energy and SOC amplitude
present for the 2DEG in this system make it promising for the consideration in transport and
optical experiments where the disorder, collision and thermal broadening prevent the
application of conventional semiconductors. It should be noted that the discussed
properties of the band structure for Bi/Si 2DEG are obtained in the framework of one
specific model with a set of parameters chosen for the best fit to experimental data.
Thus, it may be modified in the future when more insight will be gained on the
properties of Bi/Si or other similar compounds. Still, we shall see below that the
qualitative and topologically described features of the electron states studied within
this model are robust against the significant variations of the model parameters,
which is an indication of certain intrinsic and stable properties of the system.

\subsection{Edge states}

We now turn our attention to the construction of the model for the edge states localized at the opposite edges of the finite strip formed
in the 2DEG. We can start with the strip geometry where the electrons are confined along the $y$ direction in the strip $-L/2 \le y \le L/2$ and
with conventional assumption of the hard-wall boundary conditions $\Psi(x,y=\pm L/2)=0$.\cite{Koenig,Zhou,Krueckl2011,Linder,Lu}

First, the spectrum of edge states can be found by solving the eigenstate problem with the requirement of exponential dependence across the strip
direction $Oy$. It can be done by starting from the bulk Hamiltonian and replacing the quasimomentum component in the direction of confinement
by the pure imaginary variable describing the inverse localization depth corresponding to the localized states $~ \exp (\pm \Lambda y)$ which means
in our case the substitution $k_y \to -i \Lambda$.
It should be mentioned that in general $\Lambda$ can be complex with imaginary part corresponding to the oscillations of the edge
wavefunctions on top of the exponential decay\cite{Koenig,Linder} while in other models \cite{Zhou} $\Lambda$ is taken purely real,
like in our system. The reason for purely real $\Lambda$ in our model of edge states is the narrow bulk gap
formed in the bulk spectrum originating from Hamiltonian (\ref{h0}) with strong Rashba SOC. If one adds more real nonzero wavevector
components by adding the imaginary part to $\Lambda$, then the resulting energy increase will push the edge states out of the bulk gap,
making them unsuitable for the TI phase.

The eigenfunctions of this Hamiltonian can be constructed in the same nearly free-electron approximation as the bulk states (\ref{bulkwf}),
and have the following form:

\begin{eqnarray}
\Phi_{k_x \Lambda}(x,y)=e^{\Lambda y} F_{k_x \Lambda}(x), \\
F_{k_x \Lambda}(x)=\sum_n a_{n} (k_x,\Lambda) \phi_{n k_x \Lambda}(x).
\label{edgestate}
\end{eqnarray}

The spinors $\phi_{n k_x \Lambda}(x)$ can be obtained from (\ref{rf}) with the
substitution  $k_y \to -i \Lambda$, which results in a real number under the Arg function,
giving us

\begin{equation}
\phi_{n k_x \Lambda}(x)=\frac{e^{i (k_x+nG) x}}{\sqrt{2}}
\left(
\begin{array}{c}
1 \\
(\mp i) {\rm sign} (k_x+nG+\Lambda)
\end{array}
\right).
\label{edgespinor}
\end{equation}

The summation in (\ref{edgestate}) is over the 1D lattice in the reciprocal space which corresponds not to
2D hexagonal but the 1D simple lattice along the $Ox$ direction  formed by the vectors ${\bf G}_1$ and ${\bf G}_4$ from
Fig.\ref{figbulk}a, with the real space period period $a=2\pi/G$ where $G=1.08$ $\AA^{-1}$ is
the length of the ${\bf G}_i$ vector. The state (\ref{edgestate}) remains to be Bloch
function along $Ox$ with the conventional translational property $\Phi_{k_x \Lambda}(x+a,y)=e^{i k_x a} \Phi_{k_x \Lambda}(x,y)$,
while along the confinement direction the wavefunctions are exponents $\exp (\pm \Lambda y)$.
If we solve the Schr{\" o}dinger equation for our model of 2DEG at Bi/Si interface with the substitution $k_y \to -i \Lambda$
 (without considering the specific boundary conditions at this stage),
the spectrum of edge states will be obtained as a function of two parameters $(k_x, \Lambda)$. If there are eigenstates with energies
corresponding to the gap in the bulk spectrum, one can be interested in them as in potential candidates for the edge states with topological protection.

The wavefunction satisfying the boundary conditions on a single edge $\Psi_{k_x}(x,y=\pm L/2)=0$ and having
the specific energy $E=E(k_x,\Lambda)$ in the bulk gap has the form of linear combination of (\ref{edgestate}) with different
localization lengths $\Lambda_{1,2}$ corresponding to the given energy $E=E(k_x,\Lambda_{1,2})$:

\begin{equation}
\Psi_{k_x}(x,y)=\sum_{\Lambda} c_{\Lambda} \Phi_{k_x \Lambda}(x,y).
\label{edgewf}
\end{equation}

\begin{figure}[tbp]
\centering
\includegraphics*[width=85mm]{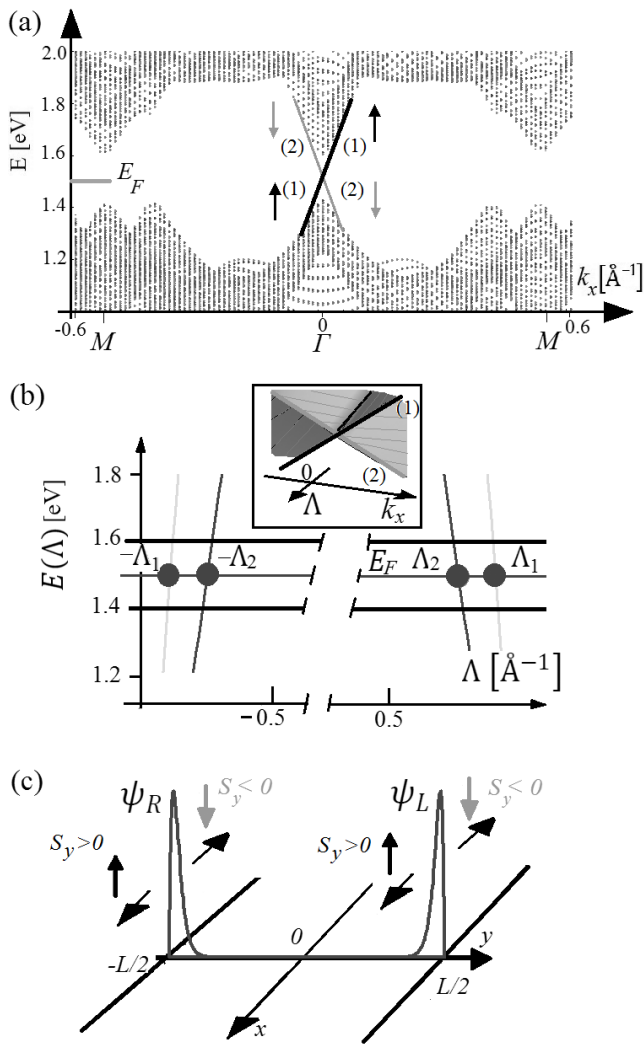}
\caption{(a) Bulk (grayscale dots on the background) and edge (black and gray linear dispersions marked as (1) and (2))
energy bands in Bi/Si strip shown vs $k_x$ for all $k_y$ (bulk states)
and for typical value of the inverse localization length $\Lambda=0.6$ $\rm{\AA}^{-1}$ (for edge states).
The edge states are formed in the bulk energy gap where the position of the Fermi level is shown,
and have the opposite spin polarization coupled to their group velocity for left-moving electrons (band (2))
and right-moving electrons (band (1)).
(b) Edge state energy dependence on the inverse localization length parameter $\Lambda$
taken for $k_x=0.05$ $\rm{\AA}^{-1}$. For the given position of the Fermi level in the bulk gap there are
two roots $\pm \Lambda_{1,2}$ for each edge of the strip giving two edge wavefunctions belonging to the corresponding branches of energy spectrum.
The inset shows the side view of the edge state spectrum vs both $k_x$ and $\Lambda$, demonstrating the 3D structure of two
branches (1) and (2) of spin split states intersecting along the line $k_x=0$.
(c) Edge states localized on the opposite borders of the strip for the energy of the edge state
equal to the Fermi level inside the bulk gap $E_F=1.5$ eV and for the strip width $L=10$ nm.
The edge states are well-localized at the corresponding edge
of the strip. For each edge there are two states propagating to the positive and negative directions
of the $Ox$ axis and having opposite spin polarizations $S_y$.}
\label{figedgespec}
\end{figure}

If the edge is represented by another and more smoothly rising potential differing from the hard wall, or more sophisticated
boundary condition is chosen, the edge wavefunction is expected to be modified mainly in the small vicinity of the edge where a  decaying tail can be formed.
Since this modification would not affect seriously the global shape of the edge state and the main localization parameter, as well as the primary
property of their possible topological stability induced by the presence of the topological invariant for the bulk states,we shall proceed with the simple
hard wall condition introduced above.
We have found that in our model the edge states indeed exist in the gap of the bulk spectrum, and their dependence on
the $k_x$ parameter is shown in Fig.\ref{figedgespec}(a) as two linear dispersion curves (1) and (2) crossing the bulk gap
for typical value of inverse localization length $\Lambda=0.6$ $\rm{\AA}^{-1}$,
together with the bulk spectrum plotted vs $k_x$ for all values of $k_y$ in the 2D Brillouin zone shown in Fig.\ref{figbulk}a.
The joint dependence on both $k_x$ and $\Lambda$ for the energy of the edge states is shown as 3D plot in the inset for Fig.\ref{figedgespec}b.
The two spin-resolved branches of energy $E=E(k_x,\Lambda)$ are shaded differently (branch (1) is dark gray and
branch (2) is light gray) there with respect to their spin projection $S_y$.
In Fig.\ref{figedgespec}(a) we also plot the mean values of the only nonvanishing spin component $S_y=\langle \Psi \mid \sigma_y \mid \Psi \rangle$
for the edge states which are coupled to their direction of motion along the strip.
The spin-up states are moving to the right with the group velocity $v_x=\frac{1}{\hbar} \frac{\partial \varepsilon}{\partial k_x}>0$
while the spin-down states are moving to the left ($v_x<0$), shown in the same grayscale level with corresponding
branches of the edge spectrum.

An important feature of the edge state spectrum is the presence of two roots $\Lambda_{1,2}$ for each value of energy for a given $k_x$,
i.e. the equation $E(k_x,\Lambda)=E_0$ has two pairs of solutions $\pm \Lambda_{1,2}$ for the left and right edge of the strip,
respectively, as it is shown in Fig.\ref{figedgespec}b for the dependence of the energy on $\Lambda$ at $k_x=0.05$ $\rm{\AA}^{-1}$ .
Such structure of energy eigenvalues is the direct consequence of the relative proximity of two branches of Rashba spectrum present
in the basis for the Hamiltonian which can be also seen for the bulk spectrum in Fig.{\ref{figbulk}}b.
This feature allows one to construct the edge states satisfying the boundary
conditions for the band of energies located in the bulk energy gap, as it is done in various models of edge states in TI.\cite{Koenig,Zhou,Linder,Lu}

The specific boundary conditions are applied to the general form of the edge state (\ref{edgewf}).
The two wavefunctions for the given energy $E=E(k_x,\Lambda_{1,2})$ satisfying the boundary condition $\Psi_L(x,y=L/2)=0$
on the left edge of the strip $y=L/2$ (when facing in the forward direction of the $Ox$ axis) and decaying into the
strip have the form (\ref{edgewf}) and can be constructed explicitly by the following superposition of the states (\ref{edgestate}):

\begin{equation}
\Psi^{(1)}_L (x,y)=F_{k_x \Lambda_1} (x) \left( e^{\Lambda_1 y} - e^{ (\Lambda_1-\Lambda_2)\frac{L}{2}+ \Lambda_2 y} \right),
\label{leftwf1}
\end{equation}

\begin{equation}
\Psi^{(2)}_L (x,y)=F_{k_x \Lambda_2} (x) \left( e^{\Lambda_2 y} - e^{ (\Lambda_2-\Lambda_1)\frac{L}{2}+ \Lambda_1 y} \right)
\label{leftwf2}
\end{equation}

where the normalization condition is implied in $F_{k_x \Lambda}(x)$.
Correspondingly, the localized wavefunctions for the right edge $y=-L/2$ can be written as

\begin{equation}
\Psi^{(1)}_R (x,y)=F_{k_x -\Lambda_1} (x) \left( e^{-\Lambda_1 y} - e^{ (\Lambda_1-\Lambda_2)\frac{L}{2}- \Lambda_2 y} \right),
\label{rightwf1}
\end{equation}

\begin{equation}
\Psi^{(2)}_R (x,y)=F_{k_x -\Lambda_2} (x) \left( e^{-\Lambda_2 y} - e^{ (\Lambda_2-\Lambda_1)\frac{L}{2}- \Lambda_1 y} \right)
\label{rightwf2}
\end{equation}

All the edge states (\ref{leftwf1}),(\ref{leftwf2}) and (\ref{rightwf1}),(\ref{rightwf2}) have
different spinor parts $F_{k_x \pm \Lambda_{1,2}}$ due to the different value of parameter $\pm \Lambda_{1,2}$, and in general may describe different spin polarization.
It should be noted that their spin properties are described by the mean value of spin calculated for the edge state
which itself is not labeled by the spin quantum number, as typical in the systems with SOC where the spin operator
does not commute with the Hamiltonian. Still, since for the 1D edge states their direction of propagation is strongly coupled
to the sign of the mean spin polarization and the edge subband index, one can link these two properties together and
describe the edge states as having a definite spin mean value.

An example of the edge wavefunctions is shown in Fig.\ref{figedgespec}(c) for the energy of the edge state equal to the Fermi level inside the bulk gap $E=E_F=1.5$ eV
and for the strip width $L=10$ nm. One can see that the edge states are well-localized at the corresponding edge of the strip on the length
of about 1 nm if the corresponding bulk topological invariant
is nontrivial, as it is discussed in the next section. The arrows indicate the direction of propagation and the spin $S_y$ of each state in pair (\ref{leftwf1}),(\ref{leftwf2}) and
(\ref{rightwf1}),(\ref{rightwf2}). The direction of the propagation of two chiral states on the one edge $y=-L/2$ in our model is the same as
on the other edge $y=L/2$, which can be explained by to the strong Rashba SOC in our system
leading, among other things, to the dominant polarization $S_y$ of the states moving in the $Ox$ direction. Here the non-compensated total spin $S_y$ can be accumulated
if the population of right-moving and left-moving electrons is unbalanced, for example, by
external electric field $E || Ox$, as it happens in various models with current-induced spin polarization both in conventional materials with strong SOC and in TI,\cite{KS}
which is also can be expected for the edge states shown in Fig.\ref{figedgespec}(c).

The form of spin polarization shown in Fig.\ref{figedgespec}(a),(c) creates a positive expectation about their topological stability for the charge transport against
the scattering on non-magnetic impurities which do not violate the TR symmetry, if supported by the analysis of the topological
properties of the bulk states indicating the presence of a nontrivial topological invariant. If we consider the backscattering,
then it is clear from Fig.\ref{figedgespec}(c) that the change of the propagation direction will lead to the spin flip, and such reversal
cancels the reflected waves and extinguishes the backscattering.\cite{KaneRMP,QiZhangRMP}
This is consistent with the arguments of the topological stability of such edge states as the participants of charge transport, which is
the required property of a system to become a topological insulator. We shall see below in Sec.III that our assumption about the 2DEG on
the Bi/Si interface as a possible TI is supported further by the analysis of the topological properties of 2D bulk states.

\section{Topological properties of bulk states}

It is known from the general theory that the stability of edge states is guaranteed by certain topological properties
of the bulk states. In particular, the system can be a TI if an integer called
$Z_2$ invariant is different from zero.\cite{KaneRMP,QiZhangRMP,KaneMele,FuKane}
 There are several ways for calculation of this invariant, and here we shall use the method proposed by Kane and Mele \cite{KaneMele}
which links the $Z_2$ index to the zeros of the Pfaffian for the interband matrix elements of the time reversal
operator between the occupied bands, which has the form $\Theta=i \sigma_y K$ for the spin-1/2 particles, where $K$ is the complex conjugation operator.
If we have only two lowest bands occupied which is the case of 2DEG in Bi/Si 2DEG, then the Pfaffian $P_{1,2}({\bf k})$ is equal
to the single off-diagonal matrix element between the Bloch functions $u_{1,2}({\bf k})$ in the occupied bands $1$ and $2$,

\begin{equation}
P_{1,2}({\bf k})=\langle u_1 ({\bf k}) \mid \Theta \mid u_2 ({\bf k}) \rangle.
\label{Pfaffian}
\end{equation}

The topological considerations provide a convenient form of using the definition (\ref{Pfaffian}) for the task of finding
new material demonstrating the properties of TI. Namely, if for the hexagonal BZ the ${\bf k}$-dependent function $P({\bf k})$ has pairs of zeros in the corners of the BZ
(or, depending on the overall symmetry, on the lines inside the BZ), then the system demonstrates the properties of TI.\cite{KaneMele,Murakami,FuKane}
There is an extensive discussion of the $Z_2$ invariant properties related to the TI including another definition of it where the matrix elements
of TR operator $\Theta$ are calculated between the states with opposite wavevectors
${\bf k}$ and $-{\bf k}$ and the TI is determined by its properties not in the entire BZ but at discrete set of "time reversal invariant points" like
$\Gamma$ or $M$ points.
The detailed discussion and all relevant mathematical connections between different approaches on the calculation of $Z_2$ invariant
can be found in the literature.\cite{KaneRMP,QiZhangRMP,FuKane,Schnyder2008,Ryu2007,Fukui2007,Roy2009,Yu2011}

\begin{figure}[tbp]
\centering
\includegraphics*[width=85mm]{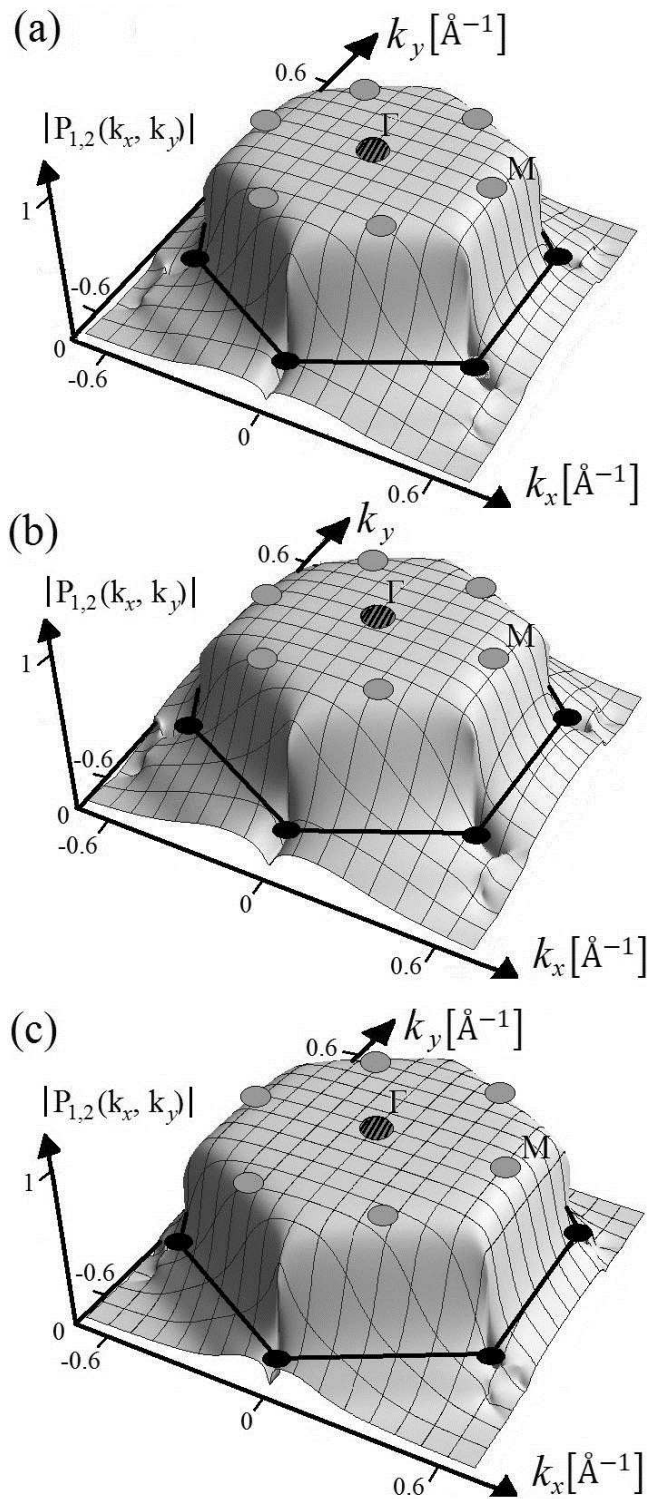}
\caption{Absolute value of the Pfaffian (\ref{Pfaffian}) in the hexagonal BZ for the 2DEG on the Bi/Si interface for different values of bulk band parameters, (a) $V_0=0.3$ eV, $\alpha_R=1.1$ $\rm{eV \cdot \AA}$;
(b) $V_0=0.6$ eV, $\alpha_R=1.1$ $\rm{eV \cdot \AA}$;
(c) $V_0=0.3$ eV, $\alpha_R=0.6$ $\rm{eV \cdot \AA}$.
The Pfaffian has three pairs of zeros at the corners of the BZ  where the visible zeros are shown as black circles while
at TR-invariant $\Gamma$ and $M$ points shown as shaded and gray circles the value  $\mid P_{1,2}\mid=1$.
These properties indicate that the $Z_2$ invariant is odd, and the topological insulator phase is present for
all three sets of material parameters.}
\label{figpfaff0}
\end{figure}

In Fig.\ref{figpfaff0}a we plot the absolute value of the Pfaffian (\ref{Pfaffian}) in the hexagonal BZ for the 2DEG on the Bi/Si interface
for the same basic set of model parameters as were used for calculations of the bulk spectrum in Fig.\ref{figbulk}b.
In order to see the possible changes in the $Z_2$ index with the variation of the system parameters, in
 Fig.\ref{figpfaff0}b,c we plot $\mid P_{1,2}\mid$ for two other sets of parameters:
in Fig.\ref{figpfaff0}b the amplitude of the periodic potential is increased compared
to the basic case shown in Fig.\ref{figbulk}b, $V_0=0.6$ eV and the amplitude of Rashba SOC is the same, $\alpha_R=1.1$
$\rm{eV \cdot\AA}$.
In Fig.\ref{figpfaff0}c the periodic potential amplitude is the same as in Fig.\ref{figpfaff0}a, $V_0=0.3$ eV, but the Rashba coupling amplitude is decreased,
$\alpha_R=0.6$ $\rm{eV \cdot \AA}$.
It is clearly seen for all cases that the Pfaffian has zeros in the corners of the BZ where the visible zeros are shown as black circles which border is shown
schematically, while  $\mid P_{1,2}\mid=1$ in the TR-invariant $\Gamma$ and $M$ points shown as
shaded and light gray circles, respectively.
There are three pairs of zeros for $\mid P_{1,2}\mid$ which indicates that the $Z_2$ invariant is odd, thus classifying the 2DEG at the Bi/Si
interface as a topological insulator with the protected edge states.\cite{KaneRMP,QiZhangRMP,KaneMele,FuKane}
The presentation of the structure of Pfaffian in Fig.\ref{figpfaff0} in the whole BZ is useful
in determining the areas where the states of different bands belong to the "even" or
"odd" subspace relatively to the action of the time reversal operator $\Theta$, in accordance with the classification proposed by Kane and Mele.\cite{KaneMele}
In our case shown in Fig.\ref{figpfaff0} we see that the major part of the BZ corresponds to the states belonging to the even subspace with $|P({\bf k})|=1$,
however by approaching the borders of the BZ the value of $|P({\bf k})|$ is modified significantly, and in the corners we arrive to the odd subspace
where $|P({\bf k})|=0$, so the property of TI is present.

One can also see in Fig.\ref{figpfaff0} that the variations of material parameters do not change significantly the topological properties of the Pfaffians
which all have the same qualitative features with $\mid P_{1,2}\mid=1$ at the TR-invariant $\Gamma$ and $M$ points and with three pairs
of zeros for $\mid P_{1,2}\mid$ in the corners of the BZ. The depth of the parameter variation present in three parts of Fig.\ref{figpfaff0} is
rather big and reaches $50\%$ which covers a wide range of possible experimental and technological fabrication of the 2DEG at the Bi/Si interface.
Still, the absolute values of Pfaffians shown in Fig.\ref{figpfaff0} look very similar to each other which indicates their qualitative topological nature
being the key for discovering new examples of topological insulators. The method of mutual analysis of chiral edge states and topological bulk properties
used in our calculations can be applied to other materials and structures.

\section{Conclusions}

We have derived a model for the one-dimensional edge states for the electrons on the bismuth on silicon interface in the finite strip
geometry. Based on the bulk nearly free-electron model, their energy dispersion was obtained inside the bulk gap being linear
in the quasimomentum. The spin polarization of edge states is linked to the direction of propagation along
the given edge which provides topological stability of these chiral modes.
The topological stability of edge states was confirmed by the structure of the interband matrix element
for the time reversal operator which was shown to be stable against the large variations of material parameters. The results of the paper may be of
interest both for the development of the topological insulator theory by providing a novel example of the material belonging to this class,
and also for the development of new spintronics and nanoelectronics devices with stable transport and operating at room temperature.

\section*{Acknowledgements}
The authors are grateful to V.Ya Demikhovskii, A.M. Satanin, A.P. Protogenov, G.M. Maximova, V.A. Burdov and A.A. Konakov for helpful discussions.
The work is supported by the Russian Foundation for Basic Research (Grants No. 13-02-00717a, 13-02-00784a).

\end{document}